\pdfoutput=1
\documentclass[twocolumn,prl,superscriptaddress,longbibliography]{revtex4-1}
\usepackage{graphicx,amsmath,amssymb}
\usepackage[colorlinks=true, citecolor=blue, urlcolor=blue, linkcolor=red]{hyperref}
\renewcommand{\section}[1]{{\par\it #1.---}\ignorespaces}

\DeclareMathOperator{\var}{Var}
\DeclareMathOperator{\mi}{min}
\DeclareMathOperator{\tr}{Tr}
\newcommand{\ii}{{i\mkern3mu}}

\newcommand{\aven}{\mathcal{N}}
\newcommand{\avens}{\mathcal{M}}
\newcommand{\braket}[1]{\langle#1\rangle}
\newcommand{\bra}[1]{\langle #1|}
\newcommand{\ket}[1]{|#1\rangle}

\newcommand{\oH}{\hat{H}}

\newcommand{\oSz}{\hat{S}^z}

\newcommand{\oSp}{\hat{S}^+}
\newcommand{\oSm}{\hat{S}^-}
\newcommand{\osig}[1][i]{\hat{\sigma}_{#1}}

\newcommand{\ovS}{\hat{\mathbf{S}}}

\newcommand{\soD}{\mathcal{\check{\mathcal{D}}}}

\begin{document}
\title{Generating Stable Spin Squeezing by Squeezed-Reservoir Engineering }
\author{Si-Yuan Bai}
\affiliation{Lanzhou Center for Theoretical Physics, Key Laboratory of Theoretical Physics of Gansu Province, Lanzhou University, Lanzhou 730000, China}
\author{Jun-Hong An}\email{anjhong@lzu.edu.cn}
\affiliation{Lanzhou Center for Theoretical Physics, Key Laboratory of Theoretical Physics of Gansu Province, Lanzhou University, Lanzhou 730000, China}
\date{\today}

\begin{abstract}
As a genuine many-body entanglement, spin squeezing (SS) can be used to realize the highly precise measurement beyond the limit constrained by classical physics. Its generation has attracted much attention recently. It was reported that $N$ two-level systems (TLSs) located near a one-dimensional waveguide can generate a SS by using the mediation effect of the waveguide. However, a coherent driving on each TLS is used to stabilize the SS, which raises a high requirement for experiments. We here propose a scheme to generate stable SS resorting to neither the spin-spin coupling nor the coherent driving on the TLSs. Incorporating the mediation role of the common waveguide and the technique of squeezed-reservoir engineering, our scheme exhibits the advantages over previous ones in the scaling relation of the SS parameter with the number of the TLSs. The long-range correlation feature of the generated SS along the waveguide in our scheme may endow it with certain superiority in quantum sensing, e.g., improving the sensing efficiency of spatially unidentified weak magnetic fields.
\end{abstract}

\maketitle

\section{Introduction}\label{sec:introduction}
A quantum-science revolution is in the making. It is expected to bring a lot of profound impacts to technological innovations. A distinguished example is quantum metrology or sensing \cite{RevModPhys.89.035002,RevModPhys.90.035005}. It pursues the development of measurement protocols with higher precision of physical quantities than the limit constrained by classical physics, i.e., the shot-noise limit, by using quantum resources. As a kind of many-body entanglement \cite{PhysRevA.46.R6797,PhysRevA.47.5138,PhysRevLett.112.155304,MA201189}, spin squeezing (SS) is one such resources. It has exhibited a wonderful power in beating the shot-noise limit \cite{bohnet2014,Bohnet1297,Hosten1552,hosten2016,Luo620,Zou6381,Bao2020}, with promising applications in quantum gyroscope \cite{PhysRevLett.98.030407,PhysRevLett.114.063002}, atomic clocks \cite{PhysRevLett.104.250801,PhysRevLett.112.190403,Komar2014,PhysRevLett.117.143004,PedrozoPenafiel2020a}, magnetometers \cite{PhysRevLett.109.253605,PhysRevLett.113.103004}, and gravimetry \cite{PhysRevLett.125.100402}. Its efficient generation is a prerequisite for further applications. The widely used method exploits the coherent spin-spin coupling in the one- and two-axis twisting models \cite{bohnet2014,Bohnet1297,Hosten1552,hosten2016,PhysRevA.47.5138,PhysRevLett.107.013601}. However, it is dynamically transient and experiences a degradation under the realistic decoherence \cite{PhysRevA.66.022314,PhysRevA.81.022106,XUE20131328}. Other methods via atom-photon couplings \cite{PhysRevLett.118.083604,PhysRevLett.122.103601,Qin2020,PhysRevLett.125.203601} and quantum nondemolition measurements \cite{PhysRevLett.85.1594,Appel10960,PhysRevA.103.023318} have also been proposed.

Waveguide quantum electrodynamics (QED) refers to a scenario where arrays of quantum emitters are coupled to a waveguide \cite{PhysRevLett.101.113903,van_Loo1494,Petersen67,Mitsch2014,Sipahigil847,Kannan2020,Kannaneabb8780,Kannan2020,PhysRevX.11.011015}. It allows for long-range interactions among the quantum emitters mediated by photons in the waveguide that is particularly interesting for quantum network applications \cite{PhysRevLett.120.213603}. Several schemes on dissipative preparation of the SS \cite{PhysRevLett.107.080503,PhysRevA.87.033831,PhysRevLett.106.020501,PhysRevLett.110.080502,PhysRevLett.115.163603,PhysRevLett.110.120402,Song2017,PhysRevA.101.042313} have been proposed based on the idea of reservoir engineering \cite{Kienzler2014,PhysRevX.5.021025,PhysRevA.91.052707,PhysRevB.94.214115,PhysRevA.98.043615} to the waveguide modes. The SS generated in such a method exists in steady states and does not depend on initial states, which endows it with the features of a long lifetime and robustness. However, a coherent driving laser on each quantum emitter is needed to stabilize the SS in these schemes. It dramatically increases the experimental difficulties when a huge number of quantum emitters are involved.

In this Letter, we propose a scheme to deterministically generate a stable SS of $N$ distant quantum emitters formed by two-level systems (TLSs) in a waveguide QED system without resorting to a coherent driving on each TLS. The main idea is based on the combined action of the technique of squeezed-reservoir engineering, which is widely used in quantum state preparation \cite{PhysRevLett.112.030602,PhysRevA.92.062311,PhysRevX.7.021041,PhysRevLett.119.023602,PhysRevLett.120.093601,PhysRevLett.126.020402}, and the mediation role of the waveguide. The waveguide enables us to manipulate the phase difference between the reservoir-induced long-range coherent and incoherent couplings of TLSs such that an effective collective spin mode of TLSs is induced by precisely controlling the positions of TLSs. Then, acting as a mold, the squeezed reservoir imprints its squeezing feature to the steady state of the collective spin. It is found that the Wineland SS parameter $\xi_R^2$, as a characterization of the improvement of sensitivity in Ramsey spectroscopy \cite{PhysRevA.46.R6797}, for our generated SS scales with the TLS number $N$ as $1.64N^{-0.54}$, which beats the ones in the one- and two-axis twisting \cite{PhysRevA.66.022314,PhysRevA.81.022106,XUE20131328} and Heisenberg \cite{PhysRevLett.110.257204} models with the realistic dissipation considered. It implies the superiority of the SS of our scheme in quantum sensing. The spatial separation of the TLSs along the waveguide in our scheme also is helpful in improving the sensing efficiency of a weak field via effectively increasing the contact area.

\begin{figure}
\centering
\includegraphics[width=.9\columnwidth]{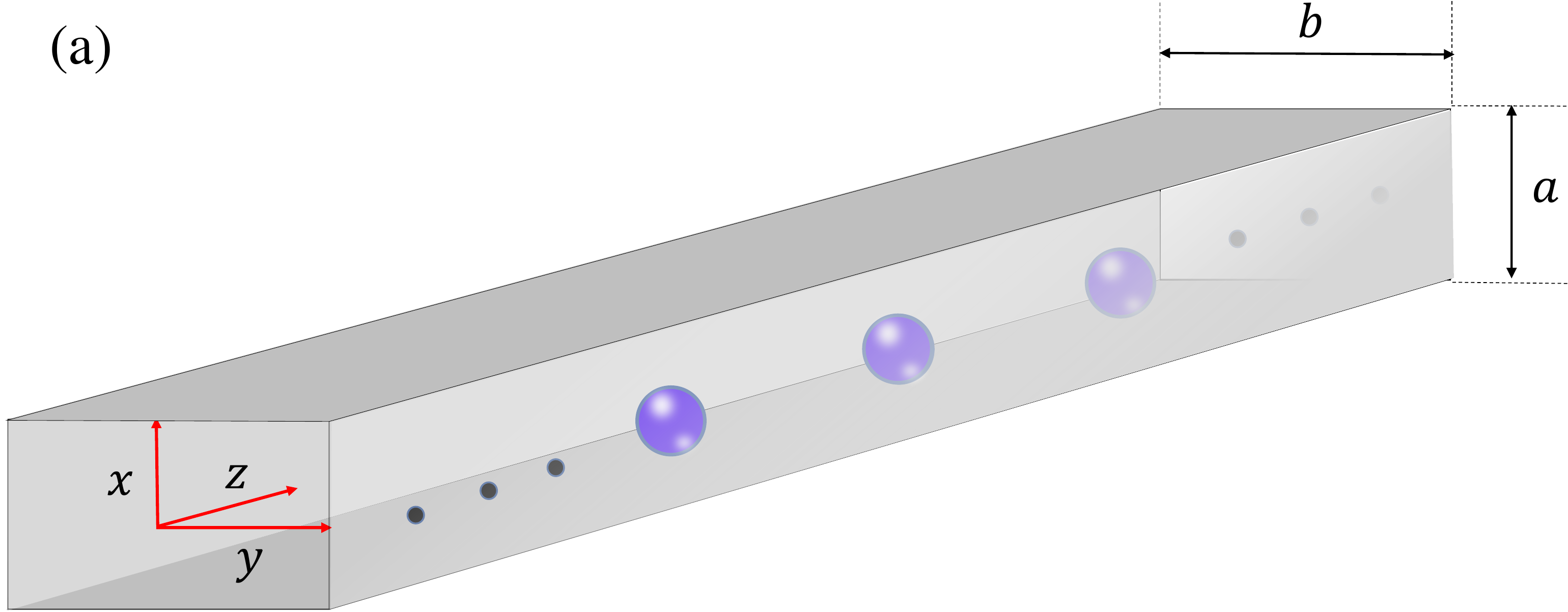}\\
\vspace{0.1cm}
\includegraphics[width=\columnwidth]{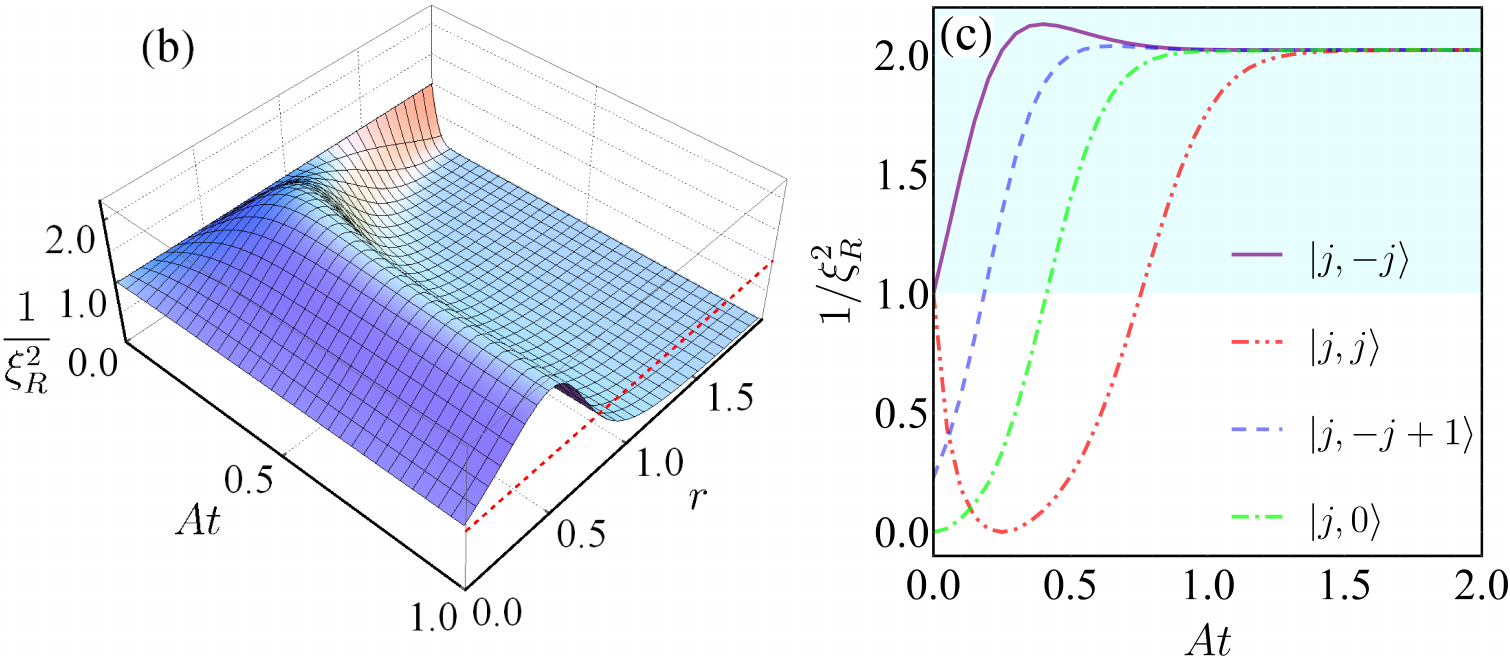}
\caption{(a) Schematics of $N$ TLSs along $z$ axis in a one-dimensional waveguide, which is driven by a broadband squeezed field.  Evolution of the inverse of the SS parameter of the TLSs in different $r$ when the initial state is $|j,-j\rangle$ with $j=N/2$ (b) and in different initial states when $r=0.5$ (c). The red dashed line in (b) denotes $\xi_R^2=1$. Other parameters are $N=10$, $\omega_0=\Delta=1.0A$, and $\alpha=0.5$. }\label{fig:dynamics}
\end{figure}

\section{System and dynamics}\label{sec:system}
We consider an array of $N$ TLSs coupled to a common electromagnetic field in a waveguide [see Fig. \ref{fig:dynamics}(a)]. The TLSs can be superconductor qubits \cite{Liu2016}, nitrogen vacancy centers \cite{Song2017}, or natural atoms \cite{Kri2018}. The waveguide may be a coupled cavity array \cite{PhysRevA.93.062122}, a metal-dielectric surface plasmon \cite{Fang2015,PhysRevResearch.1.023027}, or a photonic crystal \cite{DUTTA201641}. Its Hamiltonian is $\hat{H}=\hat{H}_\text{S}+\hat{H}_\text{R}+\hat{H}_\text{I}$ with ($\hbar=1$)
\begin{eqnarray}
\hat{H}_\text{S}&=&\sum_{i=1}^{N}\omega_{0}\hat{\sigma}_{i}^\dag\hat{\sigma}_{i},~\hat{H}_\text{R}=\sum_{k}\omega_{k}\hat{a}_{k}^{\dagger}\hat{a}_{k},\\
\hat{H}_\text{I}&=&\sum_{k,i}(g_{ki}\hat{a}_{k}+g_{ki}^{*}\hat{a}_{k}^{\dagger})(\hat{\sigma}_{i}^\dag+\hat{\sigma}_{i}),
\end{eqnarray}
where $\osig$ is the transition operator from the excited state $|e\rangle$ to the ground state $|g\rangle$ of the $i$th TLS at $\mathbf{r}_i$, with transition frequency $\omega_0$, and $\hat{a}_k$ is the annihilation operator of the $k$th field mode with frequency $\omega_k$. The coupling strength is $g_{ki}=\sqrt{\omega_k/(2\epsilon_0)}\mathbf{d}_i\cdot\mathbf{u}_k({\bf r}_i)$, where $\epsilon_0$ is the vacuum permittivity, ${\bf d}_i$ is the dipole moment of the $i$th TLS, and ${\bf u}_k ({\bf r}_i)$ is the spatial function of the $k$th mode.

We use the technique of squeezed-reservoir engineering \cite{PhysRevA.92.062311,PhysRevX.7.021041,PhysRevLett.119.023602,PhysRevLett.120.093601,PhysRevLett.126.020402} to generate the SS of TLSs. It is realized by feeding the waveguide a broadband squeezed field, which can be implemented via an optical parametric down conversion \cite{PhysRevLett.117.110801,Serikawa:16} or Josephson parametric amplification \cite{PhysRevA.86.013814,Macklin307}. The waveguide under the driving of the squeezed field acts as a squeezed vacuum reservoir. The initial state is $\rho_\text{T}(0)=\rho(0)\otimes \prod_k\hat{S}_k|0_k\rangle\langle 0_k|\hat{S}_k^\dag$, where $\hat{S}_k=\exp[r_k(e^{-i\alpha_k}\hat{a}_{k}^{2}-e^{i\alpha_k}\hat{a}_{k}^{\dag2})/2]$ with $r_k$ and $\alpha_k$ being the squeezing strength and angle, and $\ket{0_k}$ is the $k$th-mode vacuum state \cite{PhysRevLett.112.030602}. Both $r_k$ and $\alpha_k$ relate to the amplitude of the pump field, the second-order nonlinearity, and the length of nonlinear material in parametric amplification. The master equation of the TLSs under the Born-Markovian and secular approximations is \cite{book_open,PhysRevA.88.052111,SupplementalMaterial}
\begin{eqnarray}
\dot{\tilde{\rho}}(t)&=&-\ii[\Delta_\aven\sum_{i}\osig^\dag\osig+\oH_\text{DD},\tilde{\rho}(t)]\nonumber\\%
&&  +\sum_{i,j}\big\{\gamma_{ij}^-/2\big[\aven \soD_{\osig^\dag,\osig[j]}+(\aven+1)\soD_{\osig,\osig[j]^\dag}\big]\nonumber\\%
&&-\gamma_{ij}^+/2\big[\avens \soD_{\osig^\dag,\osig[j]^\dag} +\avens^*\soD_{ \osig,\osig[j]}\big]\big\}\tilde{\rho}(t),\label{eq:markov_dy_full}
\end{eqnarray}%
where $\tilde{\rho}(t)=e^{i\hat{H}_\text{S}t}\rho(t)e^{-i\hat{H}_\text{S}t}$ is the density matrix of TLSs in the interaction picture, $\aven\equiv\sinh^{2}r$ with $r\equiv r(\omega_0)$, $\avens\equiv\sqrt{\aven^2+\aven}e^{i\alpha}$ with $\alpha\equiv\alpha(\omega_0)$, and $\soD_{\hat{A},\hat{B}}\tilde{\rho} \equiv 2 \hat{A}\tilde{\rho} \hat{B}-\tilde{\rho} \hat{B}\hat{A}-\hat{B}\hat{A}\tilde{\rho}$. %
The first term of Eq. (\ref{eq:markov_dy_full}) is the reservoir-induced coherent dynamics, where $H_\text{DD}=-\sum_{i\neq j}\Delta_{ij}(\osig^\dag\osig[j]+\osig\osig[j]^\dag)/2$ is the dipole-dipole interaction of the TLSs. The interaction strengths are $\Delta_{ij}=\Delta_{ij}^{-}+\Delta_{ij}^{+}$ with $\Delta_{ij}^{\pm}=\mathcal{P}\int_{0}^{\infty}d\omega{G^-_{ij}(\omega)/(\omega\pm\omega_0)}$,
where $\mathcal{P}$ denotes the Cauchy principal value and $G^-_{ij}(\omega)\equiv\sum_k g_{ki} g_{kj}^*\delta(\omega-\omega_k)$ are the correlated spectral densities.
The frequency shift equals to $\Delta_\aven\equiv(2\aven+1)\Delta$ with $\Delta=(\Delta_{ii}^+-\Delta_{ii}^-)$, which is independent of the positions of TLSs. The second and third lines of Eq. (\ref{eq:markov_dy_full}) are the incoherent dissipation and squeezing with rates $\gamma_{ij}^\pm=2\pi G_{ij}^\pm(\omega_0)$, where $ G_{ij}^+(\omega)\equiv \sum_k g_{ki} g_{kj}\delta(\omega-\omega_k)$. It is interesting to see that the squeezed vacuum reservoir, as a common medium of the TLSs, can not only induce individual dissipation, squeezing, and frequency shift to each TLSs, but also induce correlated dissipation, squeezing, and coherent dipole-dipole interactions to the TLSs by exchanging the photons in the waveguide. It gives us sufficient room to generate a long-range correlation of TLSs via engineering the reservoir in the waveguide.

\begin{figure*}
\centering
\includegraphics[width=0.95\textwidth]{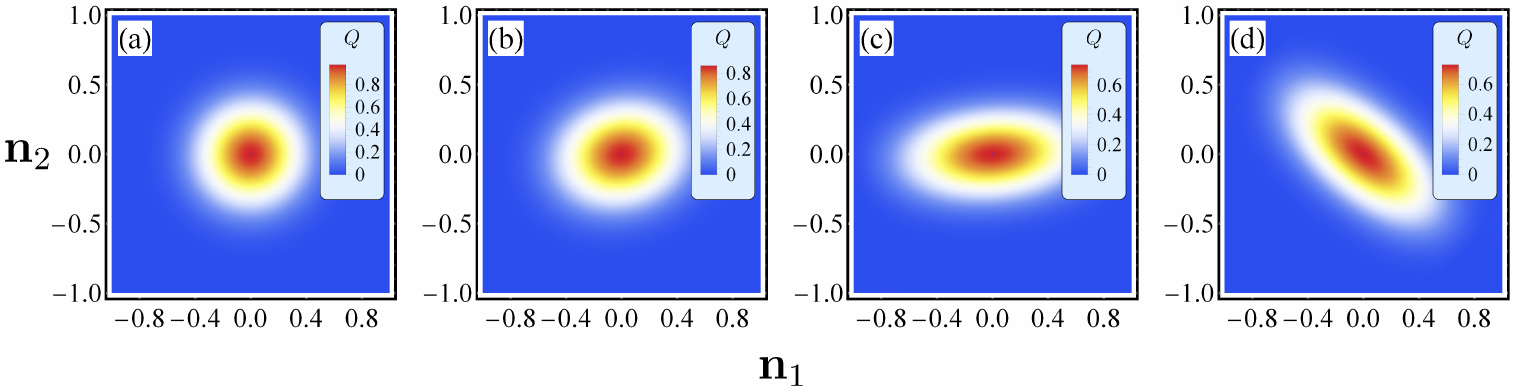}
\caption{Evolution of the Husimi's $Q$ function in the $\mathbf{n}_\perp$ plane for the initial state $|j,-j\rangle$ when $t=0$ (a), $0.01A^{-1}$ (b), $0.1A^{-1}$ (c), and $t=4.0A^{-1}$ (d). We use $N=30$, $r=0.8$, and others are the same as Fig. \ref{fig:dynamics}.}\label{fig:q}
\end{figure*}

The surface plasmonic as a waveguide experiences severe loss due to the metal absorption \cite{Fang2015}. The photonic crystal as a waveguide reflects light
only in a certain narrow frequency range \cite{DUTTA201641}.  We consider that the waveguide is formed by a rectangular hollow metal due to its high $Q$ factor \cite{Pucci2013} and wide permissable bandwidth. Its allowable electromagnetic modes are the transverse modes TE$_{mn}$ and TM$_{mn}$ above the cutoff frequency $\omega _{mn}=c[( m\pi /a) ^{2}+(n\pi /b) ^{2}]^{1/2}$, where $a$, $b$ are the transverse lengths of the waveguide, and $m$, $n$ are nonnegative integers. Their dispersion relations are $\omega ^{mn}_{k}=[( ck) ^{2}+\omega_{mn}^{2}]^{1/2}$ with $k$ being the longitudinal wave number and $c$ being the speed of light \cite{book_elect}. Assuming that the TLSs are polarized in the $z$ direction ($\mathbf{d}_i=d\mathbf{e}_{z}$), we have \cite{PhysRevA.87.033831}
\begin{equation}
G^\pm_{ij}( \omega )=\sum_{mn}\frac{\Gamma _{mn}}{2\pi }\frac{\cos [ k( z_{i}\pm z_{j}) ] }{[(\omega /\omega _{mn}) ^{2}-1]^{1/2}}\Theta ( \omega -\omega_{mn})
\end{equation} with $\Theta (x) $ as the Heaviside step function, $\Gamma_{mn}={4\omega _{mn}\tilde{u}_{mn,i}\tilde{u}_{mn,j}\over\epsilon _{0} cab}$, and $\tilde{u}_{mn,i}=d\sin ( \frac{m\pi }{a}x_{i})\sin ( \frac{n\pi }{b}y_{i}) $. Considering the dominated mode with $m=n=1$ and $\omega_0>\omega_{11}$, we can derive \cite{SupplementalMaterial}
\begin{eqnarray}
	\Delta_{ij}&=&-[\Gamma_{11}\zeta\omega_{11}/(2c)]\sin({|z_i-z_j|/ \zeta}),\label{dipl}\\
\gamma^{\pm}_{ij}&=&(\Gamma_{11}\zeta\omega_{11}/c)\cos({|z_i\pm z_j|/ \zeta}),\label{dissp}
\end{eqnarray}
where $\zeta=c(\omega_0^2-\omega_{11}^2)^{-1/2}$. Equations \eqref{dipl} and \eqref{dissp} originate from the interference of the $N$ individual interaction channels of the TLSs with the common reservoir \cite{PhysRevA.93.062122}. Remarkably, the waveguide as the reservoir medium enables us to modulate the phase difference of Eqs. \eqref{dipl} and \eqref{dissp} via tailoring the position $z_{i}$. It allows the switch-on-off for either of the two couplings and offers an opportunity for engineering the multipartite quantum correlation of the TLSs. By positioning the TLSs such that $|z_{i}\pm z_j|=2n_\pm\pi\zeta$ ($n_\pm\in \mathbb{Z}$), we have $\Delta_{ij}=0$, $\gamma^\pm_{ij}=\Gamma_{11}\zeta\omega_{11}/c\equiv A$, and Eq. (\ref{eq:markov_dy_full}) reduced to \cite{SupplementalMaterial}
\begin{align}\label{eq:markov_dy_reduced}
\dot{\tilde{\rho}}(t)=&-\ii\Delta_\mathcal{N}[ \oSz,\tilde{\rho}(t)]+\frac{A}{2}\Big[\aven \soD_{\oSp,\oSm}+(\aven+1)\soD_{\oSm,\oSp}\nonumber\\
&-\avens \soD_{\oSp,\oSp} -\avens^*\soD_{ \oSm,\oSm}\Big]\tilde{\rho}(t),
\end{align}
where $\oSz\equiv\sum_{i}(\hat{\sigma}_i^\dag\hat{\sigma}_i-1/2)$ and $\hat{S}^-=(\hat{S}^+)^\dag=\sum_i\osig$. Thus, a collective spin mode of the TLSs is induced to interplay with the common squeezed reservoir by the constructive interference among the interaction channels.

\section{Stable spin squeezing}\label{sec:steady state}
An SS is featured with a reduced quantum fluctuation in certain spin component. Defining a mean direction $\mathbf{n}=\text{Tr}(\hat{\mathbf{S}}\rho)/|\text{Tr}(\hat{\mathbf{S}}\rho)|\equiv\left( \sin \theta _{0}\cos \varphi _{0},\sin \theta_{0}\sin \varphi _{0},\cos \theta _{0}\right)$, a spin state $\rho$ is squeezed if its minimal variance in the $\mathbf{n}_\perp$ plane spanned by $\mathbf{n}_{1}=\left( \cos \theta _{0}\cos \varphi_{0},\cos\theta _{0}\sin \varphi _{0},-\sin \theta _{0}\right)$ and $\mathbf{n}_{2}=\left( -\sin \varphi _{0},\cos \varphi _{0},0\right)$ is smaller than that of the spin coherent state, i.e. $N/4$ \cite{PhysRevA.47.5138}. It is quantified by the SS parameter \cite{PhysRevA.46.R6797}
$\xi_R^2={N[\var(\ovS^\bot)]_{\mi}}\big{/}{|\text{Tr}(\hat{\mathbf{S}}\rho)|^2}$, where $\ovS^{\bot}$ is the spin in the $\mathbf{n}_\perp$ plane, $\var(\hat{O})=\braket{\hat{O}^2}-\braket{\hat{O}}^2$, and the superscript $\mi$ means the minimum in all directions. Exhibiting multipartite entanglement, the state is squeezed if $\xi_R^2<1$ \cite{PhysRevA.79.042334}. Another way to visually depict the SS is the Husimi's $Q$ function $Q( \theta ,\varphi ) =(2j+1)/(4\pi)\left\langle \theta,\varphi\right\vert \rho\left\vert \theta ,\varphi \right\rangle$ \cite{PhysRevA.47.R2460}, where $\left\vert \theta ,\varphi\right\rangle =( 1+|\eta |^{2})^{-j}e^{-\eta^* \hat{S}^-}|j,j\rangle$ with $|j,m\rangle$ ($m=-j,\cdots,j$) being the eigenstates of $\{\hat{\mathbf{S}}^2,\hat{S}^z\}$ and $\eta =-\tan (\theta /2)e^{ -i\varphi}$. The $Q$ function maps $\rho$ to a quasiclassical probability distribution in the phase space defined by $\theta$ and $\varphi$.

Via numerically solving Eq.~(\ref{eq:markov_dy_reduced}), we show in Fig.~\ref{fig:dynamics}(b) the evolution of $1/\xi_R^2$ for the initial state $|j,-j\rangle$ with $j=N/2$ in different squeezing strengths $r$. It is found that a stable SS can be formed in the regime of a moderate $r$. Thanks to the constructive role played by the squeezed reservoir in the waveguide, the system spontaneously evolves to a spin squeezed state uniquely dependent on $r$. This is in sharp contrast to the previous results \cite{PhysRevLett.110.080502,PhysRevLett.115.163603,PhysRevLett.110.120402,Song2017,PhysRevA.101.042313}, where a coherent driving field is applied on each TLS to stabilize the SS. The evolution of $1/\xi_R^2$ with chosen $r$ in different initial states $|j,m\rangle$ verifies the uniqueness of the steady state [see Fig. \ref{fig:dynamics}(c)]. We plot in Fig.~\ref{fig:q} the evolution of the projections of the $Q$ function in the $\mathbf{n}_\perp$ plane for the initial state $|j,-j\rangle$. Being isotropically distributed, it shows no SS initially. With increasing time, it shrinks in one direction at the expense of expanding in its orthogonal direction. Such an SS is kept to its steady state [see Fig. ~\ref{fig:q}(d)]. It confirms that the TLSs as a collective spin are squeezed by the squeezed reservoir.

To figure out the relation between the SS and $r$, we calculate the steady state \cite{PhysRevA.41.3782,SupplementalMaterial}
\begin{equation}
\tilde{\rho}(\infty)=\sum_{m,n=-j}^{j}p_{m}p_{n}^*\braket{\phi_m|\phi_n}\ket{\psi_m}\bra{\psi_n}.\label{eq:steady_solution}
\end{equation}
Satisfying $\hat{R}_{z}|\psi_{m}\rangle  =m|\psi_{m}\rangle$ and $\hat{R}_{z}^{\dagger}|\phi_{m}\rangle =m|\phi_{m}\rangle$ with $\hat{R}_{z}=i(4|\avens|)^{-1/2}(\hat{S}^{+}e^{ i\frac{\alpha}{2}}\sinh r-\hat{S}^{-}e^{- i\frac{\alpha}{2}}\cosh r)$, $|\psi_{m}\rangle$ and $|\phi_{m}\rangle$ forms a complete set of biorthogonal basis. The recurrence relation of coefficient $p_{m}$ is $p_{m+1}={Am-i\Delta_{\mathcal{N}}\over A(m+1)+i\Delta_{\mathcal{N}}}p_{m}$, which can be fixed by $\tr[\tilde{\rho}(\infty)]=1$.

\begin{figure}
\centering
\includegraphics[width=\columnwidth]{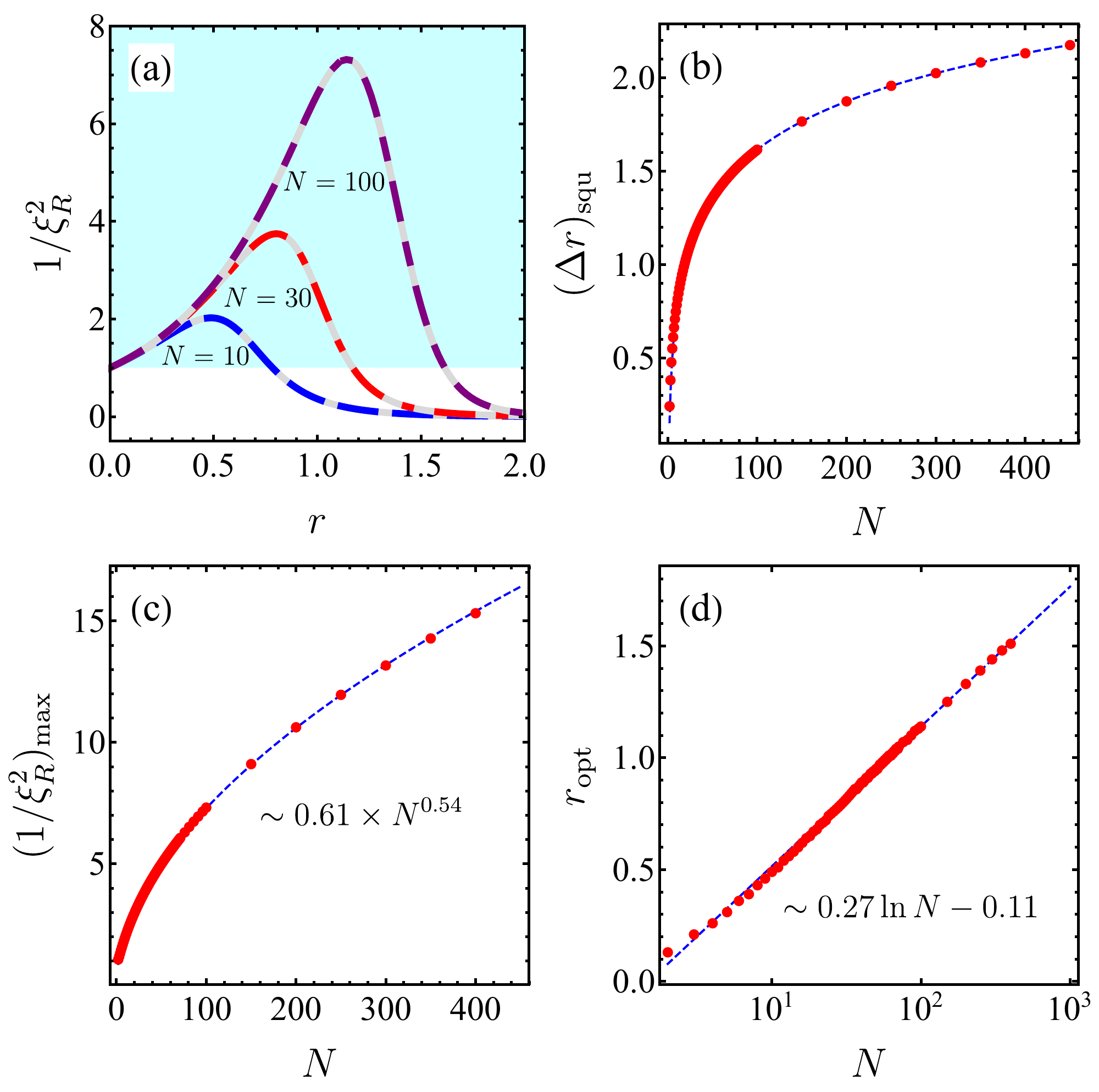}
  \caption{(a) Comparison of $1/\xi_R^2$ from the analytical solution \eqref{eq:steady_solution} (purple dot dashed, red dashed, and blue dotted lines) with the one by numerically solving Eq. \eqref{eq:markov_dy_reduced} at $t=200A^{-1}$ (gray lines) in different $N$. Range of $r$ supporting stable SS (b), maxima of $1/\xi_R^2$ (c), and optimal $r$ (d) as a function of $N$. The numerical fitting gives $(\xi_R^2)_\text{Max}=1.64N^{-0.54}$ and $r_\text{opt}=0.27\ln N-0.1$. Parameters are the same as Fig. \ref{fig:dynamics}. }\label{fig:steady}
\end{figure}

We plot in Fig.~\ref{fig:steady}(a) the comparison of the steady-state squeezing obtained by the analytical solution \eqref{eq:steady_solution} with the numerical one. It verifies the correctness of Eq.  \eqref{eq:steady_solution} in describing the steady state of Eq. \eqref{eq:markov_dy_reduced}. One observation from Fig.~\ref{fig:steady}(a) is that, with increasing $N$, the range of $r$ supporting the stable SS becomes wider and wider. Thus, the stable SS can be generated in a fairly wider parameter range if more TLSs are involved [see Fig. \ref{fig:steady}(b)]. Another feature of Fig.~\ref{fig:steady}(a) is that, even for sufficiently large $N$, the SS still tends to vanish in a gradual manner with increasing $r$. It endows our scheme with a difference from the previous ones based on the driven-dissipative Dicke model \cite{PhysRevLett.110.080502,Song2017}, where a nonequilibrium phase transition manifested by an abrupt disappearance of the SS is presented. It is understandable based on the fact that our SS is generated via purely incoherent interactions of TLSs mediated by the reservoir, while theirs is via the combined actions of the incoherent interactions and coherent driving.

The SS parameter $\xi_{R}^{2}$ describes the improvement of the sensitivity to measure the atomic frequency in Ramsey spectroscopy \cite{PhysRevA.46.R6797}. Therefore, it itself is an important quantity to characterize quantum superiority in quantum metrology \cite{Gross_2012,T_th_2014,Giovannetti2011,PhysRevLett.113.250401}. The analytical solution \eqref{eq:steady_solution} permits us to investigate the SS in the large-$N$ limit, where the numerical calculation is hard. The relation between the optimal SS and the number $N$ of the TLSs is calculated via Eq. \eqref{eq:steady_solution} [see red dots in Fig. \ref{fig:steady}(b)]. Via numerical fitting, we obtain its scaling relation as
\begin{equation}
(\xi_R^2)_\text{min}=1.64\times N^{-0.54}.
\end{equation}
According to the definition of $\xi_{R}^{2}$ \cite{PhysRevA.46.R6797}, the metrology error using this state in Ramsey spectroscopy outperforms the shot-noise limit achieved using the spin coherent state by a factor of $\sqrt{1.64}N^{-0.27}$. It is better than the previous schemes. Explicitly, the SS generated via the one-axis twisting scales as $\xi_{R}^{2}\propto N^{-2/3}$ and via two-axis twisting as $N^{-1}$ in the ideal situation \cite{PhysRevA.47.5138}. However, they tend to $N^{-1/2}$ at optimized time and to be divergent in the long-time limit when the dissipation is considered \cite{PhysRevA.66.022314,PhysRevA.81.022106,XUE20131328}. Our SS is also better than the ones in the ground state of Lipkin-Meshkov-Glick model scaling as $\xi_{R}^{2}\propto N^{-1/3}$ \cite{PhysRevLett.93.237204} and in the steady state of dissipative Heisenberg model scaling as $\xi_{R}^{2}=1/2$ \cite{PhysRevLett.110.257204}. Figure \ref{fig:steady}(c) shows the squeezing strength to achieve the best SS in different $N$. The numerical fitting reveals $r_{\text{opt}}=0.27\ln N-0.11$ \cite{SupplementalMaterial}, which gently depends on $N$. Thus, we do not bother to sharply increase $r$ to generate the SS for a large number of TLSs. This gives a useful guideline for experiments to optimize the working condition.

\section{Effect of position imperfection}
\begin{figure}
\centering
\includegraphics[width=0.95\columnwidth]{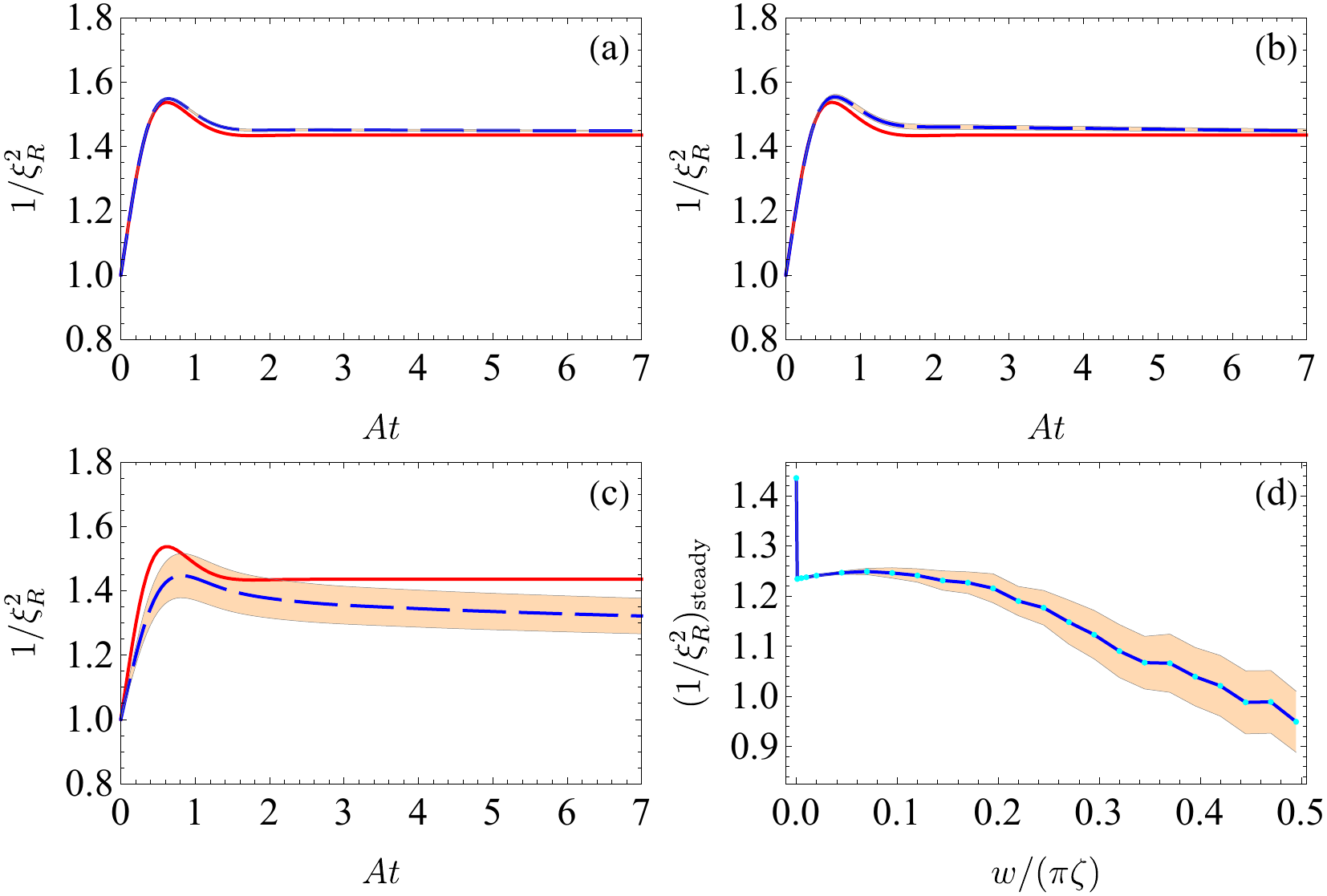}
\caption{Evolution of $1/\xi_R^2$ in the ideal (red line) and disordered (blue dashed line) cases with the standard deviations marked in orange when $w=0.03\pi\zeta$ (a), $0.06\pi\zeta$ (b), and $0.3\pi\zeta$ (c) by averaging over $100$ random configurations. (d) Steady-state $1/\xi_R^2$ in different $w$ by averaging over $200$ random configurations. $N=5$ and others are the same as Fig. \ref{fig:dynamics}.} \label{disorder}
\end{figure}

Consider that the position $z_i$ of $i$th TLS has a disorder $w\chi_i$, where $\chi_i$ is a random number uniformly distributed in $[-1,1]$ and $w$ is disorder strength. The disorder makes the collective spin mode in Eq. \eqref{eq:markov_dy_reduced} not exist anymore. Solving the original master equation \eqref{eq:markov_dy_full}, we plot in Figs. \ref{disorder}(a)- \ref{disorder}(c) the evolution of $1/\xi_R^2$ in different $w$. It can be found that the disorder introduces a decay factor roughly in a timescale $\zeta/(Aw)$ to $1/\xi_R^2$ \cite{SupplementalMaterial}. When $w$ is small, almost no observable influence can be found in a sufficiently wide time window. With increasing $w$, the decay becomes obvious. By setting the left-hand side of Eq. \eqref{eq:markov_dy_full} to zero, we plot in Fig. \ref{disorder}(d) the steady-state $1/\xi_R^2$ in different $w$. It shows that the stable SS is preserved until the disorder strength is as high as $0.4\pi\zeta$. This reveals the robustness of our scheme to the position imperfection of the TLSs.

\section{Discussion and conclusion}\label{sec:conclusion}
The current experimental advances in nanophotonics and circuit QED provide a support to our scheme \cite{Petersen67,Mitsch2014,van_Loo1494,Sipahigil847,Kannan2020}. The transmission-line waveguide mediated couplings of two 18.6mm separated TLSs have been observed \cite{van_Loo1494}. A 15 dB squeezed light corresponding to $r=1.7$ has been realized \cite{PhysRevLett.117.110801}. A 20 dB squeezing for microwave with bandwidth up to a few GHz has been realized by the Josephson parametric amplifier \cite{Macklin307}, which fulfills our requirement. The precise positioning of TLSs with 20 nm accuracy was reported \cite{Mohan2010,PhysRevB.101.205403}. Our scheme is also realizable in SiV centers as TLSs in a diamond waveguide \cite{PhysRevApplied.7.064021,PhysRevB.94.214115,PhysRevLett.120.213603}. Comparing with the SS in an atom ensemble \cite{bohnet2014,Bohnet1297,Hosten1552,hosten2016,Luo620,Zou6381,Bao2020}, our SS shows a long-range correlation. It hopefully is useful in developing quantum sensing in extremal conditions, e.g., improving the sensing efficiency of a spatially unidentified weak magnetic field via effectively increasing the contact area.

In summary, we have proposed a scheme to generate stable SS of $N$ distant TLSs in a waveguide QED system by squeezed-reservoir engineering. A collective effect of the far separated TLSs is efficiently created by the mediation role of the common squeezed reservoir in the waveguide via well-positioned TLSs. It makes the TLSs spontaneously evolve from any initial state to a spin squeezed state in the long-time limit. Our analysis reveals that the generated SS scales with the number of TLSs as $N^{-0.54}$, which outperforms the two-axis twisting and Heisenberg models with the realistic dissipation considered. Without resorting to the coherent driving on each TLS, our scheme reduces the difficulty of experiment realization in previous schemes. The recent advancement of the waveguide QED experiments indicates that our scheme is within the present experimental state of the art.

\section{Acknowledgments}\label{sec:acknowledgments}
This work is supported by the National Natural Science Foundation (Grants No. 11875150, No. 11834005, and No. 12047501).

\bibliography{bib}
\end{document}

\title{Supplemental material for ``Generating stable spin squeezing by squeezed-reservoir engineering'' }
\author{Si-Yuan Bai}
\affiliation{Lanzhou Center for Theoretical Physics, Key Laboratory of Theoretical Physics of Gansu Province, Lanzhou University, Lanzhou 730000, China}
\author{Jun-Hong An}\email{anjhong@lzu.edu.cn}
\affiliation{Lanzhou Center for Theoretical Physics, Key Laboratory of Theoretical Physics of Gansu Province, Lanzhou University, Lanzhou 730000, China}
\maketitle

\section{Derivation of the master equation}

Consider $N$ distant two-level systems (TLSs) interacting with a common squeezed reservoir confined in one-dimensional waveguide. Its Hamiltonian reads $\hat{H}=\hat{H}_{\text{S}}+\hat{H}_{\text{R}}+\hat{H}_{\text{I}}$ with ($\hbar =1$)
\begin{equation}
\hat{H}_{\text{S}}=\sum_{i=1}^{N}\omega _{0}\hat{\sigma}_{i}^{\dag }\hat{\sigma}_{i},~\hat{H}_{\text{R}}=\sum_{k}\omega _{k}\hat{a}_{k}^{\dagger }\hat{a}_{k},~\hat{H}_{\text{I}}=\sum_{k,i}(g_{ki}\hat{a}_{k}+g_{ki}^{\ast }\hat{a}_{k}^{\dagger })(\hat{\sigma}_{i}+\hat{\sigma}_{i}^{\dag }).
\end{equation}
Starting from the Liouville equation satisfied by the total system and tracing out the degrees of freedom of the reservoir, we obtain the master equation of the TLSs in the interaction picture under the Born-Markovian approximation \cite{book_open} as
\begin{equation}
\dot{\tilde{\rho}}(t)=-\int_{0}^{\infty }d\tau \text{Tr}_{\text{R}}\{[\hat{\tilde{H}}_{\text{I}}(t),[\hat{\tilde{H}}_{\text{I}}(t-\tau ),\tilde{\rho}(t)\otimes R_{0}]]\},
\end{equation}
where $\tilde{\rho}(t)=e^{i\hat{H}_{\text{S}}t}\rho (t)e^{-i\hat{H}_{\text{S}}t}$, $\hat{\tilde{H}}_{\text{I}}(t)=e^{i(\hat{H}_{\text{S}}+\hat{H}_{\text{R}})t}\hat{H}_{\text{I}}e^{-i(\hat{H}_{\text{S}}+\hat{H}_{\text{R}})t}$, and $R_{0}=\prod_{k}\hat{S}_{k}|0_{k}\rangle \langle 0_{k}|\hat{S}_{k}^{\dag }$, with $\hat{S}_{k}=\exp [r_k(e^{-i\alpha_k}\hat{a}_{k}^{2}-e^{i\alpha_k} \hat{a}_{k}^{\dag
2})/2]$, is the initial state of the squeezed reservoir. Introducing the notation $\hat{\tilde{H}}_{\text{I}}(t)\equiv \sum_{i}\hat{s}_{i}(t)\hat{\Gamma}_{i}(t)$ with $\hat{s}_{i}(t)=\hat{\sigma}_{i}^{\dag }e^{i\omega _{0}t}+\hat{\sigma}_{i}e^{-i\omega _{0}t}$ and $\hat{\Gamma}_{i}(t)=\sum_{k}(g_{ki}\hat{a}_{k}e^{-i\omega _{k}t}+g_{ki}^{\ast }\hat{a}_{k}^{\dagger }e^{i\omega _{k}t})$, we obtain
\begin{equation}
\dot{\tilde{\rho}}(t)=-\sum_{i,j}\int_{0}^{\infty }d\tau \{[\hat{s}_{i}(t)\hat{s}_{j}(t-\tau )\tilde{\rho}(t)-\hat{s}_{j}(t-\tau )\tilde{\rho}(t)\hat{s}_{i}(t)]\text{Tr}_{\text{R}}[\hat{\Gamma}_{i}(t)\hat{\Gamma}_{j}(t-\tau)R_{0}]+\text{h.c.}\}.  \label{smmast}
\end{equation}
The correlation functions read
\begin{equation}
\text{Tr}_{\text{R}}[\hat{\Gamma}_{i}(t)\hat{\Gamma}_{j}(t-\tau)R_{0}]=\int_{0}^{\infty }d\omega \{G_{ij}^{-}(\omega )[e^{-i\omega \tau }(\mathcal{N}_\omega+1)+e^{i\omega \tau }\mathcal{N}_\omega]-G_{ij}^{+}(\omega )[\mathcal{M}_\omega e^{-i\omega (2t-\tau )}+\text{c.c.}]\},
\end{equation}
where $\mathcal{N}_\omega=\sinh ^{2}r(\omega)$, $\mathcal{M}_\omega=\sqrt{\mathcal{N}_\omega(\mathcal{N}_\omega+1)}e^{i\alpha(\omega) }$, and $G_{ij}^{-}(\omega )=\sum_{k}g_{ki}g_{kj}^{\ast}\delta (\omega -\omega _{k})$ and $G_{ij}^{+}(\omega)=\sum_{k}g_{ki}g_{kj}\delta (\omega -\omega _{k})$. Here we have used the fact that both of $G_{ij}^{\pm }(\omega )$ are real functions for
one-dimensional waveguide. Then Eq. (\ref{smmast}) can be rewritten as
\begin{eqnarray}
\dot{\tilde{\rho}}(t) &=&-\sum_{i,j}\{\mathcal{A}_{i,j}[(\hat{\sigma}_{i}^{\dag }\hat{\sigma}_{j}^{\dag }\tilde{\rho}(t)-\hat{\sigma}_{j}^{\dag }\tilde{\rho}(t)\hat{\sigma}_{i}^{\dag })e^{i2\omega _{0}t}+\hat{\sigma}_{i}\hat{\sigma}_{j}^{\dag }\tilde{\rho}(t)-\hat{\sigma}_{j}^{\dag }\tilde{\rho}(t)\hat{\sigma}_{i}]  \notag \\
&&+\mathcal{B}_{ij}[(\hat{\sigma}_{i}\hat{\sigma}_{j}\tilde{\rho}(t)-\hat{\sigma}_{j}\tilde{\rho}(t)\hat{\sigma}_{i})e^{-i2\omega _{0}t}+\hat{\sigma}_{i}^{\dag }\hat{\sigma}_{j}\tilde{\rho}(t)-\hat{\sigma}_{j}\tilde{\rho}(t)\hat{\sigma}_{i}^{\dag }]+\text{h.c.}\}.  \label{smmmstdd}
\end{eqnarray}
The utilization of $\int_{0}^{\infty }d\tau e^{-i(\omega \pm \omega_{0})\tau }=\pi \delta (\omega \pm \omega _{0})-i\frac{\mathcal{P}}{\omega\pm \omega _{0}}$, with
$\mathcal{P}$ denoting the Cauchy principal value, results in the defined $\mathcal{A}_{i,j}$ and $\mathcal{B}_{i,j}$ as
\begin{eqnarray}
\mathcal{A}_{i,j} &=&\int_{0}^{\infty }d\tau \int_{0}^{\infty }d\omega e^{-i\omega _{0}\tau }\{G_{ij}^{-}(\omega )[e^{-i\omega \tau }(\mathcal{N}_\omega+1)+e^{i\omega \tau}\mathcal{N}_\omega]-G_{ij}^{+}(\omega )[\mathcal{M}_\omega e^{-i\omega (2t-\tau )}+\text{c.c.}]\}\notag \\
&=&\mathcal{N}_{\omega_0}\frac{\gamma _{ij}^{-}}{2}-\mathcal{M}_{\omega_0}\frac{\gamma _{ij}^{+}}{2}e^{-i2\omega _{0}t}-i(\Delta _{ij}^{+}+\Delta _{ij}^{\mathcal{N},+})+i\Delta _{ij}^{\mathcal{N},-}-i\Delta _{ij}^{\prime\mathcal{M}, -}+i\Delta_{ij}^{\prime\mathcal{M}^{\ast } ,+}, \\
\mathcal{B}_{ij} &=&\int_{0}^{\infty }d\tau \int_{0}^{\infty }d\omega e^{i\omega _{0}\tau }\{G_{ij}^{-}(\omega )[e^{-i\omega \tau }(\mathcal{N}_\omega+1)+e^{i\omega \tau }\mathcal{N}_\omega]-G_{ij}^{+}(\omega )[\mathcal{M}_\omega e^{-i\omega (2t-\tau )}+\text{c.c.}]\}\notag \\
&=&(\mathcal{N}_{\omega_0}+1)\frac{\gamma _{ij}^{-}}{2}-\mathcal{M}_{\omega_0}^{\ast }\frac{\gamma_{ij}^{+}}{2}e^{i2\omega _{0}t}+i\Delta _{ij}^{\mathcal{N},+}-i(\Delta _{ij}^{-}+\Delta _{ij}^{\mathcal{N},-})-i(\Delta_{ij}^{\prime\mathcal{M}^{\ast } ,+})^*+i(\Delta _{ij}^{\prime\mathcal{M}, - })^*,
\end{eqnarray}
where
\begin{eqnarray}
\gamma _{ij}^{\pm }=2\pi G_{ij}^{\pm }(\omega _{0}),~~\Delta _{ij}^{\pm }=\mathcal{P}\int_{0}^{\infty }d\omega \frac{G_{ij}^{-}(\omega )}{\omega \pm\omega _{0}},~~\Delta _{ij}^{\mathcal{N},\pm }=\mathcal{P}\int_{0}^{\infty }d\omega \frac{G_{ij}^{-}(\omega )\mathcal{N}_\omega}{\omega \pm\omega _{0}},\\
\Delta _{ij}^{\prime\mathcal{M}, - }=\mathcal{P}\int_{0}^{\infty}d\omega \frac{G_{ij}^{+}(\omega )\mathcal{M}_\omega e^{- i2\omega t}}{\omega - \omega _{0}},~~\Delta_{ij}^{\prime\mathcal{M}^{\ast } ,+}=\mathcal{P}\int_{0}^{\infty}d\omega \frac{G_{ij}^{+}(\omega )\mathcal{M}^*_\omega e^{i2\omega t}}{\omega + \omega _{0}}.
\end{eqnarray}
The neglect of the rapidly oscillating terms in the master equation under the secular approximation \cite{PhysRevA.88.052111} converts Eq. (\ref{smmmstdd}) into
\begin{eqnarray}
\dot{\tilde{\rho}}(t) &=&\sum_{i,j}\{i[(\Delta _{ij}^{+}+\Delta _{ij}^{\mathcal{N},+})(\hat{\sigma}_{i}\hat{\sigma}_{j}^{\dag }\tilde{\rho}(t)-\tilde{\rho}(t)\hat{\sigma}_{j}\hat{\sigma}_{i}^{\dag })+(\Delta _{ij}^{-}+\Delta _{ij}^{\mathcal{N},-})(\hat{\sigma}_{i}^{\dag }\hat{\sigma}_{j}\tilde{\rho}(t)-\tilde{\rho}(t)\hat{\sigma}_{j}^{\dag }\hat{\sigma}_{i})] \nonumber\\
&&-i[\Delta _{ij}^{\mathcal{N},+}(\hat{\sigma}_{i}^{\dag }\hat{\sigma}_{j}\tilde{\rho}(t)-\tilde{\rho}(t)\hat{\sigma}_{j}^{\dag }\hat{\sigma}_{i})+\Delta _{ij}^{\mathcal{N},-}(\hat{\sigma}_{i}\hat{\sigma}_{j}^{\dag }\tilde{\rho}(t)-\tilde{\rho}(t)\hat{\sigma}_{j}\hat{\sigma}_{i}^{\dag })] \nonumber\\
&&+\gamma _{ij}^{-}/2[\mathcal{N}_{\omega _{0}}\mathcal{\check{D}}_{\hat{\sigma}_{i}^{\dag },\hat{\sigma}_{j}}+(\mathcal{N}_{\omega _{0}}+1)\mathcal{\check{D}}_{\hat{\sigma}_{i},\hat{\sigma}_{j}^{\dag }}]-\gamma _{ij}^{+}/2[\mathcal{M}_{\omega _{0}}\mathcal{\check{D}}_{\hat{\sigma}_{i}^{\dag },\hat{\sigma}_{j}^{\dag }}+\mathcal{M}_{\omega _{0}}^{\ast }\mathcal{\check{D}}_{\hat{\sigma}_{i},\hat{\sigma}_{j}}]\tilde{\rho}(t)\},
\end{eqnarray}
where $\soD_{\hat{A},\hat{B}}\tilde{\rho}(t) \equiv 2 \hat{A}\tilde{\rho}(t) \hat{B}-\tilde{\rho}(t) \hat{B}\hat{A}-\hat{B}\hat{A}\tilde{\rho}(t)$. According to the commutation relation $[\hat{\sigma}_{i}^{\dag },\hat{\sigma}_{j}]=\hat{\sigma}_{i}^{z}\delta _{i,j}$ and $\hat{\sigma}_{i}^{\dag }\hat{\sigma}_{i}=(1+\hat{\sigma}_{i}^{z})/2$, we have
\begin{eqnarray}
&&(\Delta _{ij}^{+}+\Delta _{ij}^{\mathcal{N},+})(\hat{\sigma}_{i}\hat{\sigma}_{j}^{\dag }\tilde{\rho}(t)-\tilde{\rho}(t)\hat{\sigma}_{j}\hat{\sigma}_{i}^{\dag })+(\Delta _{ij}^{-}+\Delta _{ij}^{\mathcal{N},-})(\hat{\sigma}_{i}^{\dag }\hat{\sigma}_{j}\tilde{\rho}(t)-\tilde{\rho}(t)\hat{\sigma}_{j}^{\dag }\hat{\sigma}_{i})\nonumber \\
&=&\frac{\Delta _{ij}^{+}+\Delta _{ij}^{\mathcal{N},+}+\Delta _{ij}^{-}+\Delta _{ij}^{\mathcal{N},-}}{2}[\hat{\sigma}_{i}^{\dag }\hat{\sigma}_{j}+\hat{\sigma}_{i}\hat{\sigma}_{j}^{\dag },\tilde{\rho}(t)]+(\Delta _{ii}^{-}+\Delta _{ii}^{\mathcal{N},-}-\Delta _{ii}^{+}-\Delta _{ii}^{\mathcal{N},+})[\hat{\sigma}_{i}^{\dag}\hat{\sigma}_{i},\tilde{\rho}(t)],\\
&&\Delta _{ij}^{\mathcal{N},+}(\hat{\sigma}_{i}^{\dag }\hat{\sigma}_{j}\tilde{\rho}(t)-\tilde{\rho}(t)\hat{\sigma}_{j}^{\dag }\hat{\sigma}_{i})+\Delta _{ij}^{\mathcal{N},-}(\hat{\sigma}_{i}\hat{\sigma}_{j}^{\dag }\tilde{\rho}(t)-\tilde{\rho}(t)\hat{\sigma}_{j}\hat{\sigma}_{i}^{\dag }) \nonumber\\&=&\frac{\Delta _{ij}^{\mathcal{N},+}+\Delta_{ij}^{\mathcal{N},-}}{2}[\hat{\sigma}_{i}^{\dag }\hat{\sigma}_{j}+\hat{\sigma}_{i}\hat{\sigma}_{j}^{\dag },\tilde{\rho}(t)]-(\Delta _{ii}^{\mathcal{N},-}-\Delta _{ii}^{\mathcal{N},+})[\hat{\sigma}_{i}^{\dag }\hat{\sigma}_{i},\tilde{\rho}(t)].
\end{eqnarray}
Finally, we arrive at
\begin{equation}
\dot{\tilde{\rho}}(t)=-i[\Delta _{\mathcal{N}}\sum_{i}\hat{\sigma}_{i}^{\dag}\hat{\sigma}_{i}+\hat{H}_{\text{DD}},\tilde{\rho}]+\sum_{i,j}\{\gamma
_{ij}^{-}/2[\mathcal{N}_{\omega _{0}}\mathcal{\check{D}}_{\hat{\sigma}_{i}^{\dag },\hat{\sigma}_{j}}+(\mathcal{N}_{\omega _{0}}+1)\mathcal{\check{D}}_{\hat{\sigma}_{i},\hat{\sigma}_{j}^{\dag }}]-\gamma _{ij}^{+}/2[\mathcal{M}_{\omega _{0}}\mathcal{\check{D}}_{\hat{\sigma}_{i}^{\dag },\hat{\sigma}_{j}^{\dag }}+\mathcal{M}_{\omega _{0}}^{\ast }\mathcal{\check{D}}_{\hat{\sigma}_{i},\hat{\sigma}_{j}}]\}\tilde{\rho}(t),  \label{smfnmst}
\end{equation}
where $\Delta _{\mathcal{N}}=2(\Delta _{ii}^{\mathcal{N},+}-\Delta_{ii}^{\mathcal{N},-})+\Delta _{ii}^{+}-\Delta_{ii}^{-}$ and $\hat{H}_{\text{DD}}=-\sum_{i,j}\Delta _{ij}(\hat{\sigma}_{i}^{\dag}\hat{\sigma}_{j}+\hat{\sigma}_{i}\hat{\sigma}_{j}^{\dag})/2$ with $\Delta _{ij}=\Delta _{ij}^{+}+\Delta _{ij}^{-}$. Rewriting the frequency shift $\Delta _{\mathcal{N}}=-2\omega_0\mathcal{P}\int_{0}^{\infty }d\omega \frac{G_{ij}^{-}(\omega )(1+2\mathcal{N}_\omega)}{\omega^2 -\omega^2 _{0}}$, we see that the integration is dominated by two poles $\omega=\pm\omega_0$. However, the pole $\omega=-\omega_0$ has no contribution because $G_{ii}^{-}(\omega )$ in the numerator
is nonzero only when $\omega>\omega_{11}$, with $\omega_{11}$ being the cutoff frequency, for our considered rectangular hollow metal waveguide. Therefore, we obtain $\Delta _{\mathcal{N}}\simeq (2\mathcal{N}_{\omega_0}+1)(\Delta_{ii}^{+}-\Delta_{ii}^{-})$.

It is interesting to see that, although the squeezing strength $r(\omega)$ and angle $\alpha(\omega)$ are mode-dependent, only the ones in the resonant mode $r(\omega_0)$ and $\alpha(\omega_0)$ with the TLSs have contribution to the reservoir-induced frequency shift, dipole-dipole interactions, and spontaneous emissions. %Without loss of generality, we consider in the main text that the squeezing strength and angle are mode-independent.

\section{Rectangular hollow metal waveguide}

\begin{figure}
\centering
\includegraphics[angle=0,width=.55\textwidth]{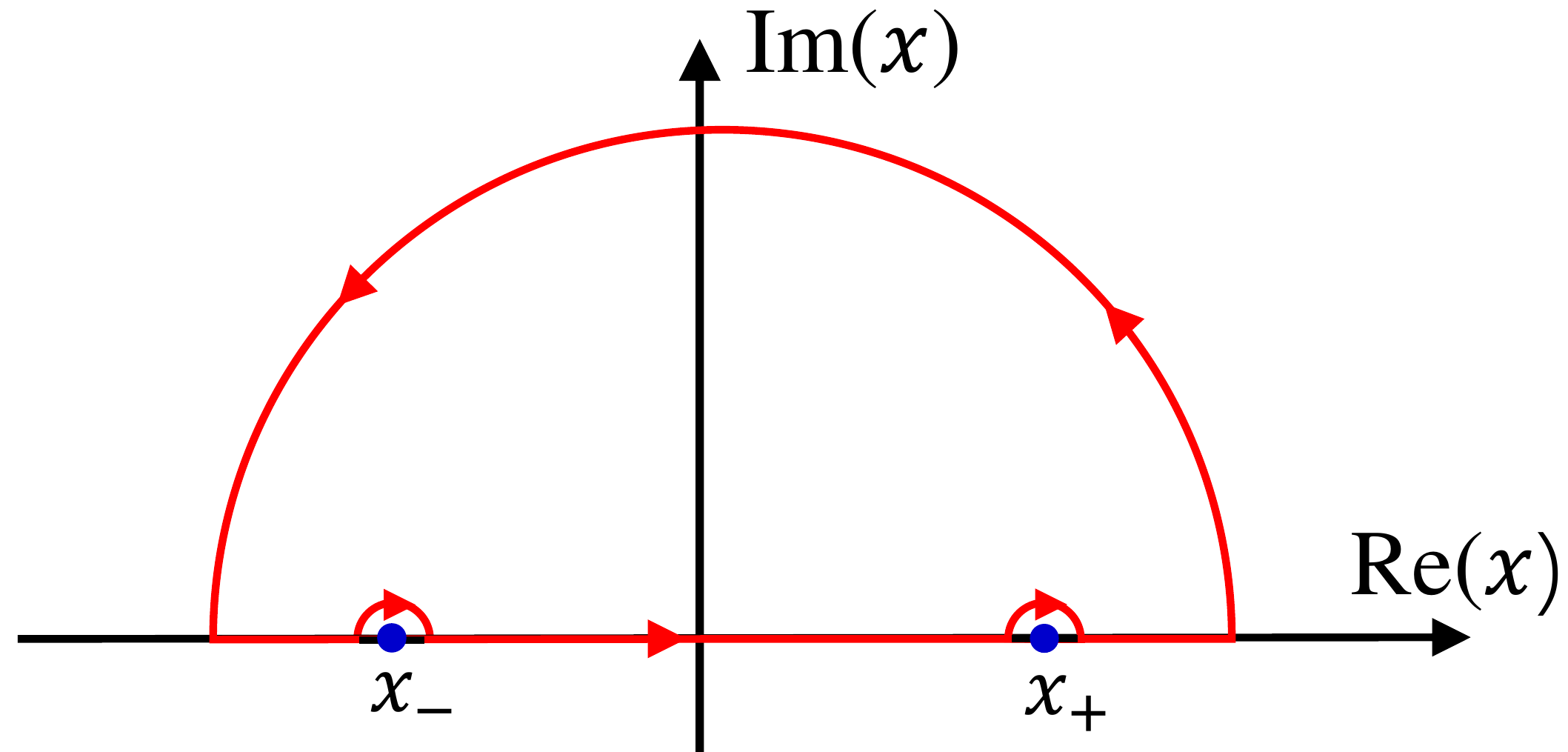}
\caption{ Path of the contour integration in the complex plane $\text{Re}(x)+i\text{Im}(x)$ for the calculation of $\Delta _{ij}$.}\label{contin}
\end{figure}

When the waveguide is formed by a rectangular hollow metal, the electromagnetic modes in the waveguide are the transverse modes TE$_{mn}$ and TM$_{mn}$ with the cutoff frequency $\omega _{mn}=c[(m\pi /a)^{2}+(n\pi/b)^{2}]^{1/2}$, where $a$, $b$ are the transverse lengths of the waveguide, and $m$, $n$ are non-negative integers. Their dispersion relations are $\omega _{k}^{mn}=[(ck)^{2}+\omega _{mn}^{2}]^{1/2}$ with $k$ being the longitudinal wave number and $c$ being the speed of light. Considering the dominate mode $m=n=1$ under the condition that the TLSs are polarized in the $z$ direction ($\mathbf{d}_{i}=d\mathbf{e}_{z}$), we can calculate the spectral densities as \cite{PhysRevA.87.033831}
\begin{equation}
G_{ij}^{\pm }(\omega )=\frac{\Gamma _{11}}{2\pi }\frac{\cos [k(z_{i}\pm z_{j})]}{[(\omega /\omega _{11})^{2}-1]^{1/2}}\Theta (\omega -\omega _{11})\label{smmspct}
\end{equation}%
with $\Theta (\omega -\omega _{11})$ being the Heaviside step function, $\Gamma _{11}=4\omega _{11}\tilde{u}_{11,i}\tilde{u}_{11,j}/(\epsilon _{0}cab)$, and $\tilde{u}_{11,i}=d\sin (\frac{\pi }{a}x_{i})\sin (\frac{\pi }{b}y_{i})$. In the following, we calculate the dipole-dipole coupling strength. The substitution of Eq. (\ref{smmspct}) into the definition of $\Delta _{ij}$, we have
\begin{equation}
\Delta _{ij}=\frac{\Gamma _{11}}{2\pi }\mathcal{P}\int_{\omega_{11}}^{+\infty }\frac{\cos (kz_{ij})}{\sqrt{(\omega /\omega _{11})^{2}-1}}\frac{1}{\omega ^{2}-\omega _{0}^{2}}d\omega =\frac{\Gamma _{11}}{2\pi }\mathcal{P}\int_{-\infty}^{+\infty }\frac{\exp (i\omega _{11}z_{ij}x/c)}{x^{2}+1-\omega _{0}^{2}/\omega _{11}^{2}}dx.  \label{smdelt}
\end{equation}
where $z_{ij}\equiv |z_{i}-z_{j}|$ and $x\equiv \sqrt{(\omega /\omega_{11})^{2}-1}$. Making an analytical extension of $x$ to a complex variable and drawing a contour to encircle the upper half complex plane (see Fig. \ref{contin}), we, according to Cauchy's residue theorem \cite{book:1123190}, obtain
\begin{equation}
\Delta _{ij}=\frac{\Gamma _{11}}{2\pi }[2\pi i\sum_{\text{Inside the contour}}\text{Res}(\text{poles})+\pi i\sum_{\text{Real axis}}\text{Res}(\text{simple poles})],
\end{equation}
where Res$(\cdot )$ denotes the residue contributed from different kinds of poles.\ Focusing on $\omega _{0}>\omega _{11}$, we can see that Eq. (\ref{smdelt}) has only two simple poles $x_{\pm }=\pm \sqrt{\omega_{0}^{2}/\omega _{11}^{2}-1}$ in real axis. Then, we have $\Delta _{ij}=\frac{i\Gamma _{11}}{2 }\sum_{i=\pm }$Res$(x_{i})$. It can be readily evaluated the residues Res$(x_{\pm })=\pm \exp (\pm i\sqrt{\omega_{0}^{2}-\omega _{11}^{2}}z_{ij}/c)/(2\sqrt{\omega _{0}^{2}/\omega
_{11}^{2}-1})$. Therefore, we have
\begin{equation}
\Delta _{ij}=-[\Gamma \zeta \omega _{11}/(2c)]\sin (z_{ij}/\zeta ),
\end{equation}%
where $\zeta =c(\omega _{0}^{2}-\omega _{11}^{2})^{-1/2}$. On the other hand, it is natural to obtain
\begin{equation}
\gamma _{ij}^{\pm }=2\pi G_{ij}^{\pm }(\omega _{0})=(\Gamma _{11}\zeta \omega _{11}/c)\cos [(z_{i}\pm z_{j})/\zeta ].
\end{equation}
It is interesting to see that we can change the phase difference between $\Delta _{ij}$ and $\gamma _{ij}^{\pm }$ to switch on or off one of them by controlling the positions of TLSs. If their positions satisfy $z_{i}\pm z_{j}=2n_{\pm }\pi \zeta $ with $n_{\pm }$ being integers, then we have $\Delta _{ij}=0$ and $\gamma _{ij}^{\pm }=\Gamma _{11}\zeta \omega_{11}/c\equiv A$. The master equation (\ref{smfnmst}) turns to
\begin{equation}
\dot{\tilde{\rho}}(t)=-i[\Delta _{\mathcal{N}}\hat{S}^{z},\tilde{\rho}(t)]+A/2[\mathcal{N\check{D}}_{\hat{S}^{+},\hat{S}^{-}}+(\mathcal{N}+1)
\mathcal{\check{D}}_{\hat{S}^{-},\hat{S}^{+}}-\mathcal{M\check{D}}_{\hat{S}^{+},\hat{S}^{+}}-\mathcal{M}^{\ast }\mathcal{\check{D}}_{\hat{S}^{-},\hat{S}^{-}}]\}\tilde{\rho}(t),  \label{smclma}
\end{equation}
where $\hat{S}^{z}\equiv \sum_{i}(\hat{\sigma}_{i}^{\dag }\hat{\sigma}_{i}-1/2)$ and $\hat{S}^{\pm }=\sum_{i}\sigma ^{\pm }$.

\section{The steady state}

\label{appst} The master equation (\ref{smclma}) can be written as
\begin{equation}
\dot{\tilde{\rho}}(t)=-i\Delta _{\mathcal{N}}[\hat{S}_{z},\tilde{\rho}(t)]+\frac{A}{2}[\mathcal{N}\mathcal{\check{\mathcal{D}}}_{\hat{\mathbb{S}}_{+},
\hat{\mathbb{S}}_{-}}+(\mathcal{N}+1)\mathcal{\check{\mathcal{D}}}_{\hat{\mathbb{S}}_{-},\hat{\mathbb{S}}_{+}}+|\mathcal{M}|(\mathcal{\check{\mathcal{
D}}}_{\hat{\mathbb{S}}_{+},\hat{\mathbb{S}}_{+}}+\mathcal{\check{\mathcal{D}}}_{\hat{\mathbb{S}}_{-},\hat{\mathbb{S}}_{-}})]\tilde{\rho}(t),\label{appmast}
\end{equation}
where $\hat{\mathbb{S}}_{\pm }=\hat{S}^{\pm }e^{\pm i\frac{\alpha +\pi}{2}}$. After introducing
\begin{eqnarray}
\hat{R}_{z} &=&(4|\mathcal{M}|)^{-1/2}(\hat{\mathbb{S}}_{+}\sinh r+\hat{\mathbb{S}}_{-}\cosh r),  \label{apprrz} \\
\hat{R}_{\pm } &=&\pm (4|\mathcal{M}|)^{-1/2}(\hat{\mathbb{S}}_{+}\sinh r-\hat{\mathbb{S}}_{-}\cosh r)-\hat{S}_{z},  \label{apprpm}
\end{eqnarray}
we can recast Eq. \eqref{appmast} into
\begin{equation}
\dot{\tilde{\rho}}(t)=i{\Delta _{\mathcal{N}}/2}[\hat{R}_{+}+\hat{R}_{-},\tilde{\rho}(t)]+2A|\mathcal{M}|\mathcal{\check{\mathcal{D}}}_{\hat{R}_{z},\hat{R}_{z}^{\dag }}\tilde{\rho}(t).  \label{mastereq}
\end{equation}
It is easy to verify the following commutation relations \cite{PhysRevA.41.3782}
\begin{eqnarray}
&&[\hat{R}_{+},\hat{R}_{-}]=2\hat{R}_{z},\quad \lbrack \hat{R}_{z},\hat{R}_{\pm }]=\pm \hat{R}_{\pm },  \label{appcommure} \\
&&[\hat{R}_{+}^{\dag },\hat{R}_{-}^{\dag }]=-2\hat{R}_{z}^{\dag },\quad\lbrack \hat{R}_{z}^{\dag },\hat{R}_{\pm }^{\dag }]=\mp \hat{R}_{\pm }^{\dag}.  \label{appcommure2}
\end{eqnarray}

The eigenvectors of the non-Hermitian operator $\hat{R}_{z}$ and its Hermitian conjugate operator $\hat{R}_{z}^\dag$, i.e., $\hat{R}_{z}|\psi_{m}\rangle =m|\psi_{m}\rangle$ and $\hat{R}_{z}^{\dagger}|\phi_{m}\rangle =m|\phi_{m}\rangle$, form a complete set of biorthogonal basis.
\begin{eqnarray}
&&\sum_{m}|\psi_{m}\rangle\langle\phi_{m}|=\sum_{m}|\phi_{m}\rangle\langle\psi_{m}|=\mathbb{I},~\langle\phi_m|\psi_n\rangle=\delta_{m,n}.~~~~~~
\end{eqnarray}
It can be calculated that
\begin{eqnarray}
&&|\psi_{m}\rangle =e^{\vartheta \hat{S}_{z}}e^{-{\frac{\pi}{4}}(\hat{\mathbb{S} }_{+}-\hat{\mathbb{S} }_{-})}|j,m\rangle.  \label{applef} \\
&&|\phi_{m}\rangle =e^{-\vartheta \hat{S}_{z}}e^{-{\frac{\pi}{4}}(\hat{\mathbb{S} }_{+}-\hat{\mathbb{S} }_{-})}|j,m\rangle,  \label{apprig}
\end{eqnarray}
with $j={N/2}$, $m=-N/2,-N/2+1,\cdots,N/2$, and $\vartheta=\ln\sqrt{\tanh r}$. According to the commutation relations \eqref{appcommure} and \eqref{appcommure2}, we have
\begin{eqnarray}
\hat{R}_{\pm}|\psi_{m}\rangle =\sqrt{(j\mp m)(j\pm m+1)}|\psi_{m\pm1}\rangle,\label{apprpmaa} \\
\hat{R}^\dag_{\mp}|\phi_{m}\rangle =\sqrt{(j\mp m)(j\pm m+1)}|\phi_{m\pm1}\rangle.  \label{appoperat}
\end{eqnarray}
We expand $\tilde{\rho}(t)$ in the biorthogonal basis as
\begin{equation}
\tilde{\rho}(t) =\sum_{m,n=-j}^{j}\rho_{mn}(t)|\psi_{m}\rangle\langle\psi_{n}|,\label{mastereq2}
\end{equation}%
with $\rho_{mn}(t)=\langle\phi_m|\tilde{\rho}(t)|\phi_n\rangle$. Rewriting $\hat{R}_{z}^{\dagger}=(4|\mathcal{M}|)^{-1/2}[\hat{R}_+-\hat{R}_-+2\cosh (2r)\hat{R}_z]$, we have
\begin{align}
\hat{R}_{z}^{\dagger}|\psi_{m}\rangle =&\frac{1}{\sqrt{4|\mathcal{M}|}}[\sqrt{(j-m)(j+m+1)}|\psi_{m+1}\rangle -\sqrt{(j+m)(j-m+1)}|\psi_{m-1}\rangle
+2m\cosh(2r)|\psi_{m}\rangle].  \label{con-3}
\end{align}
Using Eqs. \eqref{apprpmaa}, \eqref{mastereq2}, and \eqref{con-3}, we convert Eq. \eqref{mastereq} into
\begin{eqnarray}
\langle\phi_{k}|\dot{\tilde{\rho}}(t)|\phi_{k^{\prime }}\rangle&= & \Big\{Ac_{kk^{\prime }}(t)\langle\phi_{k}|\phi_{k^{\prime }}\rangle k[k^{\prime}\sinh(2r)-k\cosh(2r)]+\frac{ i\Delta_{\mathcal{N}}-A(k-1)}{2}c_{k-1,k^{\prime }}(t)\langle\phi_{k-1}|\phi_{k^{\prime }}\rangle\sqrt{(j+k)(j-k+1)}  \notag \\
&&+\frac{i\Delta_{\mathcal{N}}+A(k+1)}{2}c_{k+1,k^{\prime}}(t)\langle\phi_{k+1}|\phi_{k^{\prime }}\rangle\sqrt{(j-k)(j+k+1)} \Big\}+\text{H.c.},  \label{appmascoef}
\end{eqnarray}
where we have defined ${\rho}_{mn}(t)=c_{mn}(t)\langle\phi_m|\phi_n\rangle$. Using Eq. \eqref{appoperat}, one can prove
\begin{eqnarray}
\langle\phi_{k-1}|\phi_{k^{\prime }}\rangle\sqrt{(j+k)(j-k+1)}=\langle\phi_k|\hat{R}_+|\phi_{k^{\prime }}\rangle=[k^{\prime}\sinh(2r)-k\cosh(2r)]\langle\phi_{k}|\phi_{k^{\prime}}\rangle-\langle\phi_{k}|\hat{S}_{z}|\phi_{k^{\prime }}\rangle,\label{app1} \\
\langle\phi_{k+1}|\phi_{k^{\prime }}\rangle\sqrt{(j-k)(j+k+1)}=\langle\phi_k|\hat{R}_-|\phi_{k^{\prime }}\rangle=[k\cosh(2r)-k^{\prime
}\sinh(2r)]\langle\phi_{k}|\phi_{k^{\prime }}\rangle-\langle\phi_{k}|\hat{S}_{z}|\phi_{k^{\prime }}\rangle,  \label{app2}
\end{eqnarray}
where $\hat{R}_\pm=\mp\cosh(2r) \hat{R}_{z}\pm\sinh(2r) \hat{R}_{z}^{\dagger}-\hat{S}_{z}$ derived from Eqs. \eqref{apprrz} and \eqref{apprpm} have been used. The substitution of Eqs. \eqref{app1} and \eqref{app2} into Eq. \eqref{appmascoef} results in
\begin{eqnarray}
\langle\phi_{k}|\dot{\tilde{\rho}}(t)|\phi_{k^{\prime }}\rangle&= & \frac{\langle\phi_{k}|\phi_{k^{\prime }}\rangle }{2}[k^{\prime
}\sinh(2r)-k\cosh(2r)]\big\{\lbrack i\Delta_{\mathcal{N}}-A(k-1)]c_{k-1,k^{\prime }}(t)-[i\Delta_{\mathcal{N}}+A(k+1)]c_{k+1,k^{\prime }}(t)  \notag \\
&& +2Akc_{kk^{\prime }}(t)\big\}-\frac{\langle\phi_{k}|\hat{S}_{z}|\phi_{k^{\prime }}\rangle }{2}\big\{\lbrack i\Delta_{\mathcal{N}}-A(k-1)]c_{k-1,k^{\prime }}(t)+[i\Delta_{\mathcal{N}}+A(k+1)]c_{k+1,k^{\prime }}(t)\big\}+\text{H.c.}.~~~~~~
\end{eqnarray}
The steady-state equation requires $\langle\phi_{k}|\dot{\tilde{\rho}}(t)|\phi_{k^{\prime }}\rangle=0$. It is satisfied by
\begin{align}
[i\Delta_{\mathcal{N}}+A(k+1)]c_{k+1,k^{\prime }}= & (Ak-i\Delta_{\mathcal{N}})c_{k,k^{\prime }},~ [i\Delta_{\mathcal{N}}-A(k-1)]c_{k-1,k^{\prime }}= -
(Ak+i\Delta_{\mathcal{N}})c_{k,k^{\prime }},  \label{appsolut}
\end{align}%
where $\langle\phi_{k^{\prime }}|\hat{S}_{z}|\phi_{k}\rangle=\langle\phi_{k}|\hat{S}_{z}|\phi_{k^{\prime }}\rangle$ has been used. Due to the
independence of $k^{\prime }$ in Eq. \eqref{appsolut}, the variables in the steady-state solution $c_{k,k^{\prime }}$ can be separated as $c_{k,k^{\prime }}=p_kq_{k^{\prime }}$, where $q_k=p_k^*$. Then we obtain
\begin{align}
[A(k+1)+i\Delta_{\mathcal{N}}]p_{k+1}= & (Ak-i\Delta_{\mathcal{N}})p_{k}.
\label{apprecur}
\end{align}
The combination of the normalization condition with this recurrence relation \eqref{apprecur} can give the final result of the steady state.

\section{Scaling relation of $r_\text{opt}$}
We here prove that the scaling relation $r_\text{opt}\propto0.27\ln N$ is intrinsically related to $(1/\xi_R^2)_\text{max}\propto N^{0.54}$. Using Holstein-Primakoff transformation $\hat{S}^{z}=-N/2+\hat{a}^{\dagger}\hat{a}$ and $\hat{S}^{-}=(\hat{S}^{+})^{\dagger}=\sqrt{N}\sqrt{1-\frac{\hat{a}^{\dagger}\hat{a}}{N}}\hat{a}\approx\sqrt{N}\hat{a}$ in the large-$N$ limit, we can convert Eq. (7) to
\begin{eqnarray}
\dot{\tilde{\rho}}(t) & = & -i[\Delta_{\mathcal{N}}\hat{a}^{\dagger}\hat{a},\tilde{\rho}(t)] +\frac{AN}{2}\{\mathcal{N}\check{\mathcal{D}}_{\hat{a}^\dag,\hat{a}}+(\mathcal{N}+1)\check{\mathcal{D}}_{\hat{a},\hat{a}^\dag}-\mathcal{M}\check{\mathcal{D}}_{\hat{a},\hat{a}} -\mathcal{M}^{*}\check{\mathcal{D}}_{\hat{a}^\dag,\hat{a}^\dag}\}\tilde{\rho}(t).\label{smmst}
\end{eqnarray}
The equations of motions are readily derived from Eq. \eqref{smmst}
\begin{eqnarray}
\frac{d}{dt}\langle\hat{a}^{\dagger2}\rangle & = & -i2\Delta_{\mathcal{N}}\langle\hat{a}^{\dagger2}\rangle-AN\langle\hat{a}^{\dagger2}\rangle+AN\mathcal{M}^{*},\\
\frac{d}{dt}\langle\hat{a}^{2}\rangle & = & i2\Delta_{\mathcal{N}}\langle\hat{a}^2\rangle-AN\langle\hat{a}^2\rangle+AN\mathcal{M},\\
\frac{d}{dt}\langle\hat{a}^{\dagger}\hat{a}\rangle & = & -AN\langle\hat{a}^{\dagger}\hat{a}\rangle+AN\mathcal{N}.
\end{eqnarray}
Their steady-state solutions are
\begin{eqnarray}
\langle\hat{a}^{\dagger}\hat{a}\rangle_{\text{steady}} = \mathcal{N},~~\langle\hat{a}^{\dagger2}\rangle_{\text{steady}} & = &\langle\hat{a}^2\rangle_{\text{steady}}^*= \frac{\mathcal{M}^{*}}{i2\Delta_{\mathcal{N}}/(AN)+1}\approx\mathcal{M}^{*}.\label{ssstds}
\end{eqnarray}
Because the mean spin direction of the steady state of Eq. (7) is $\mathbf{n}=(0,0,-1)$, the $\mathbf{n}_\perp$ plane is just the $x$-$y$ plane. Using Eq. \eqref{ssstds}, we can calculate
\begin{align*}
\langle(\hat{S}_{x}\cos\theta+\hat{S}_{y}\sin\theta)^{2}\rangle_{\text{steady}} & =\frac{N}{2}\{\sqrt{\mathcal{N}(\mathcal{N}+1)}\cos(2\theta-\alpha)+(\mathcal{N}+\frac{1}{2})\},
\end{align*}
where $\hat{S}_{x}=\frac{\sqrt{N}}{2}(\hat{a}^{\dagger}+\hat{a})$
and $\hat{S}_{y}=\frac{\sqrt{N}}{2i}(\hat{a}^{\dagger}-\hat{a})$. Its minimum is achieved when $\theta=\frac{\alpha+\pi}{2}$. Then we find \begin{equation}1/\xi_{R}^{2}=[2\mathcal{N}+1-2\sqrt{\mathcal{N}(\mathcal{N}+1)}]^{-1}.
\end{equation}
In large $N$ limit, we have $1/\xi_{R}^{2}\propto4\mathcal{N}\simeq e^{2r}$. Remembering $(1/\xi_R^2)_\text{max}=0.61N^{0.54}$, we readily achieve its corresponding $r$ as $r_\text{opt}=0.27\ln N+\text{Constant}$.

\section{Effect of weak position disorder}
Consider that the position $z_i$ of $i$th TLS experiences a disorder $w\chi_i$, where $\chi_i$ is a random number uniformly distributed in $[-1,1]$ and $w$ is the disorder strength. Thus, Eqs. (5) and (6) in the main text changes into
\begin{align}
\Delta_{ij} & =-[\Gamma_{11}\zeta\omega_{11}/(2c)]\sin(|z_{i}-z_{j}+w(\chi_{i}-\chi_{j})|/\zeta),\label{smdeta}\\
\gamma_{ij}^{\pm} & =(\Gamma_{11}\zeta\omega_{11}/c)\cos(|z_{i}\pm z_{j}+w(\chi_{i}\pm\chi_{j})|/\zeta).\label{smgam}
\end{align}
The conservation law of the collective spin is broken by the disorder. The involved dimension of the Liouvellian space of the density matrix in Eq. (3) is $2^{N}\times2^N$. The effect of the disorder on the spin-squeezing generation can be evaluated only via numerical calculation. However, when the disorder strength is weak, we still can obtain some analytical result. Expanding Eqs. \eqref{smdeta} and \eqref{smgam} for the first-order $w$ and remembering $|z_{i}\pm z_j|=2n_\pm\pi\zeta$, we have $\gamma_{ij}^{\pm}={\Gamma_{11}\zeta\omega_{11}}/{c}=A$ and $
\Delta_{ij}  =-A{w|\chi_{i}-\chi_{j}|}/{(2\zeta)}$. Due to the disorder feature, such dipole-dipole interactions with rate $\Delta_{ij}$ contribute an extra dephasing effect. Therefore, keeping Lindblad terms in Eq. (7) in the main text unchanged, the weak disorder only introduces the dephasing effect. The time scale of this dephasing can be roughly estimated as $\zeta/(Aw)$. Therefore, within the considerable time duration, its influence on the dynamics of spin squeezing is extremely small [see Figs. 4(a) and 4(b) in the main text].

\bibliography{bib}